\documentclass{sigkddExp}

\usepackage{graphicx}

\usepackage{amssymb}
\usepackage{amsmath}

\usepackage{booktabs}
\usepackage{multirow}
\usepackage{enumitem}
\usepackage{color}
\usepackage{xspace}
\usepackage{adjustbox}
\usepackage{hyperref}
\usepackage{url}
\usepackage{algorithm}
\usepackage{algpseudocode}
\usepackage{array} 
\usepackage{caption}
\usepackage{subcaption}
\usepackage{tabularx}
\usepackage{tcolorbox}
\usepackage{listings}
\usepackage{pifont}
\usepackage{xcolor}
\let\citep\cite
\let\citet\cite

\definecolor{mygray}{RGB}{226, 226, 226}
\definecolor{myred}{RGB}{252, 142, 142}
\definecolor{mygreen}{RGB}{147, 255, 143}
\definecolor{myblue}{RGB}{144, 155, 255}
\definecolor{myyellow}{RGB}{253, 253, 143}
\definecolor{mypurple}{RGB}{255, 142, 250}
\definecolor{softblue}{RGB}{100, 149, 237}
\definecolor{mygreen}{RGB}{62,123,39}
\newcommand{\dataset}{\mbox{OmniRouteEval}\xspace}

\makeatletter
\renewcommand{\section}{%
  \@startsection{section}{1}{\z@}%
  {-8pt plus -2pt minus -2pt}
  {16pt}
  {\baselineskip 14pt\secfnt\@ucheadtrue}%
}
\renewcommand{\subsection}{%
  \@startsection{subsection}{2}{\z@}%
  {-8pt plus -2pt minus -2pt}
  {16pt plus 1pt minus 1pt}
  {\secfnt}%
}
\makeatother

\begin{document}
%

\title{OmniRouter: Budget and Performance \\ Controllable Multi-LLM Routing}
%

%



\author{
\noindent Kai Mei, Wujiang Xu, Minghao Guo, Shuhang Lin, Yongfeng Zhang \\
Department of Computer Science, Rutgers University
\\
\texttt{\{kai.mei, wujiang.xu, minghao.guo, shuhang.lin, yongfeng.zhang\}@rutgers.edu}
}
\maketitle
\begin{abstract}
\vspace{-10pt}
Large language models (LLMs) deliver superior performance but require substantial computational resources and operate with relatively low efficiency, while smaller models can efficiently handle simpler tasks with fewer resources. LLM routing is a crucial paradigm that dynamically selects the most suitable large language models from a pool of candidates to process diverse inputs, ensuring optimal resource utilization while maintaining response quality. Existing routing frameworks typically model this as a locally optimal decision-making problem, selecting the presumed best-fit LLM for each query individually, which overlooks global budget constraints, resulting in ineffective resource allocation.
To tackle this problem, we introduce OmniRouter, a fundamentally controllable routing framework for multi-LLM serving. Instead of making per-query greedy choices, OmniRouter models the routing task as a constrained optimization problem, assigning models that minimize total cost while ensuring the required performance level. 
Specifically, a hybrid retrieval-augmented predictor is designed to predict the capabilities and costs of LLMs. After obtaining the predicted cost and performance, we utilize a constrained optimizer for cost-optimal assignments that employs Lagrangian dual decomposition with adaptive multipliers. It iteratively converges toward the globally optimal query-model allocation, dynamically balancing latency minimization against quality thresholds while adhering to heterogeneous capacity constraints. 
Experiments show that OmniRouter achieves up to 6.30\% improvement in response accuracy while simultaneously reducing computational costs by at least 10.15\% compared to competitive router baselines. The code and the dataset are available at \url{https://github.com/dongyuanjushi/OmniRouter}.

\end{abstract}

\section{Introduction}

\begin{figure*}[tb!]
    \centering
    \includegraphics[width=0.8\linewidth]{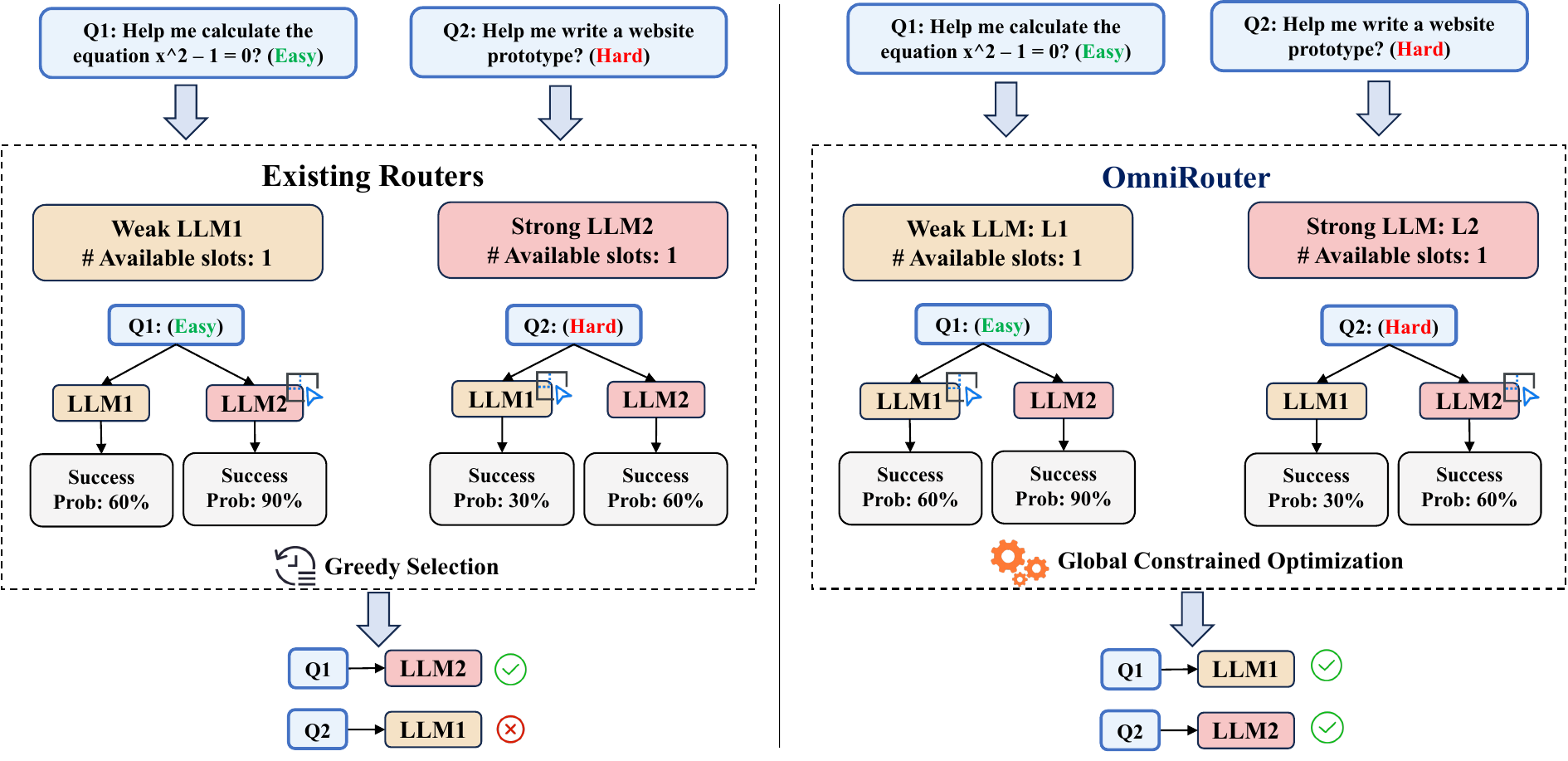}
    \vspace{-8pt}
    
    \caption{Comparison between traditional greedy routers and OmniRouter. Left: Greedy routers select models based on per-query optimization, leading to suboptimal allocations where Q1 (simple query) comes first and is assigned to the strong LLM, blocking Q2 (complex query) from accessing it. As a result, Q2 fails when assigned to the weak LLM. Right: OmniRouter employs constrained optimization to consider the global query distribution and model capabilities. It assigns the simple query to LLM1 (sufficient for the task) and reserves LLM2 for the complex query, thus maximizing overall success rate.}
    \label{fig:routing_comparison}
\end{figure*}

Large Language Models (LLMs) have demonstrated remarkable capabilities, powering a diverse range of applications from chatbots \citep{achiam2023gpt, team2023gemini, dubey2024llama, guo2025deepseek, hua2024trustagent, zhang2024ai, hua2024disentangling,iagent} and code assistants \citep{hui2024qwen2, wei2023magicoder, nijkamp2023codegen2, zhu2024deepseek} to computer-use agents. This success has spurred the widespread integration of LLMs into modern systems \citep{sun2024llumnix, mei2024aios, packer2023memgpt, shi2025from, zheng2023efficiently, kwon2023efficient,amem}. As LLM inference accelerates through hardware improvements, these LLM-integrated systems increasingly deploy not just one, but a set of LLMs with varying sizes, capabilities, and speeds as serving endpoints. This multi-LLM paradigm necessitates intelligent routing: the crucial task of directing incoming user queries to the most appropriate LLM instance to balance performance goals and resource efficiency. The design of effective LLM routers has become an active area of research. Recent works have proposed various routing strategies to optimize for cost \citep{ong2024routellm, carrot2025, hybridllm2024}, latency \citep{jin2023s3, stojkovic2024dynamollm, panda2025adaptive, zheng2024response}, and performance \citep{zhang2025capability, jin2025two, chen2024routerdc, zhuang2024embedllm}. 
Most existing LLM routing frameworks predominantly treat routing as a sequence of independent, greedy decisions. For each incoming query, these routers select a model based on a local optimization criterion (e.g., lowest predicted latency, cheapest model predicted to succeed) without considering system-wide resource limitations or overall performance targets. These approaches fundamentally fail to achieve Pareto-efficient resource allocation across diverse query workloads distributed among multiple LLMs. 
When operating under budget constraints, these greedy routers cannot globally optimize the performance-resource tradeoff, leading to suboptimal overall system efficacy. For instance, expending computational resources on marginally improving responses to simple queries may leave insufficient capacity for complex queries where performance improvements would be more valuable, which can been in \autoref{fig:routing_comparison}. More critically, these localized decision frameworks cannot effectively enforce global constraints such as maintaining target response quality levels while adhering to strict resource budgets, resulting in either performance degradation or resource over-utilization when deployed at scale.
In summary, these approaches confront a fundamental optimization dilemma: maintaining optimal overall performance while operating under computational resource constraints and stringent budgetary limitations.

Recognizing these limitations, we introduce OmniRouter, a controllable routing framework that fundamentally reimagines how query-model assignments should be determined. Unlike existing approaches, OmniRouter formalizes routing in multi-LLM systems as a \textbf{constrained optimization problem} \citep{zhang2020solving, bertsekas2014constrained, homaifar1994constrained} with global performance requirements and operational constraints and proposes a two-stage routing solution. 
Inspired by the success of retrieval-augmented generation (RAG) in reducing hallucination of LLMs, we design a retrieval-augmented predictor at the first stage to enhance the prediction-based router. At the second stage, we design the constrained optimizer which employs a dual gradient-based approach to navigate the solution space by adjusting model selections based on quality requirements and model concurrency.
Our contributions are as follows:


\noindent $\bullet$\; We propose OmniRouter, a routing framework for multi-LLM serving that fundamentally regards routing as a constrained optimization problem rather than a series of greedy decisions, enabling system-level control of both performance and budget constraints.

\noindent $\bullet$\; Extensive experiments show that OmniRouter can outperform competitive baselines across various serving scenarios and demonstrate significant improvements in response quality (up to 6.30\%), cost efficiency (at least 10.15\%), especially under tight operational constraints.

\section{Problem Formalization} \label{sec:prb}

Prior methods \citep{ong2024routellm, fu2024efficient, parkar2024selectllm} typically score or rank models for each query independently based on heuristics such as cost-effectiveness or response likelihood. While effective in simple settings, such formulations struggle to account for system-level constraints such as limited model concurrency, global quality targets, and budget bounds that are critical in real-world multi-LLM deployments. In contrast, our approach models it as a constrained optimization problem. 

Formally, given $N$ queries and $M$ models, let $a_{i,j} \in [0,1]$ denote the capability of whether the model $j$ can successfully answer query $i$, and $c_{i,j}$ denote the money cost for answering this query. Each model $j$ is subject to a concurrency limit $L_j$, and the overall system must maintain a minimum average performance constraint $\alpha$ across all $N$ queries to ensure response quality. Our objective is to assign queries to models such that the total cost is minimized, while ensuring that the overall performance exceeds $\alpha$ and that no model exceeds its concurrency constraint.
This problem can be formulated as the following: 
\begin{align} \label{eq:1}
\min_{x} \quad
& \sum_{i=1}^N \sum_{j=1}^M c_{i,j} x_{i,j} \quad s.t. \quad \frac{1}{N} \sum_{i=1}^N \sum_{j=1}^M a_{i,j} x_{i,j} \geq \alpha \\ \nonumber
& \quad \sum_{i=1}^N x_{i,j} \leq L_j, \forall j  \quad \sum_{j=1}^M x_{i,j} = 1, \forall i
\end{align}


\noindent It is a constrained optimization problem \citep{bertsekas2014constrained, homaifar1994constrained} with global constraint $\alpha$ and local constraint $L$. We will elaborate how to solve this problem in the following section.

\section{Methodology} \label{sec:method}
As the capability $a_{i,j}$ and computational cost $c_{i,j}$ for each model-query pair are inherently uncertain at routing decision time. This uncertainty makes direct optimization of the routing variable $x_{i,j}$ particularly challenging. Joint optimization couples errors between prediction and allocation \citep{feng2023towards} and make the routing struggle with adaptability when query distributions shift or new models \citep{yu2022gilbo,qin2023towards, zhang2023s}. 
To address these challenges, we propose a two-stage approach that decouples the prediction of uncertain variables: $a_{i,j}$ and $c_{i,j}$ from the optimization of routing decisions $x_{i,j}$. 

\subsection{Retrieval-augmented Predictor}
\begin{figure*}[tb!]
    \centering
    \includegraphics[width=0.85\linewidth]{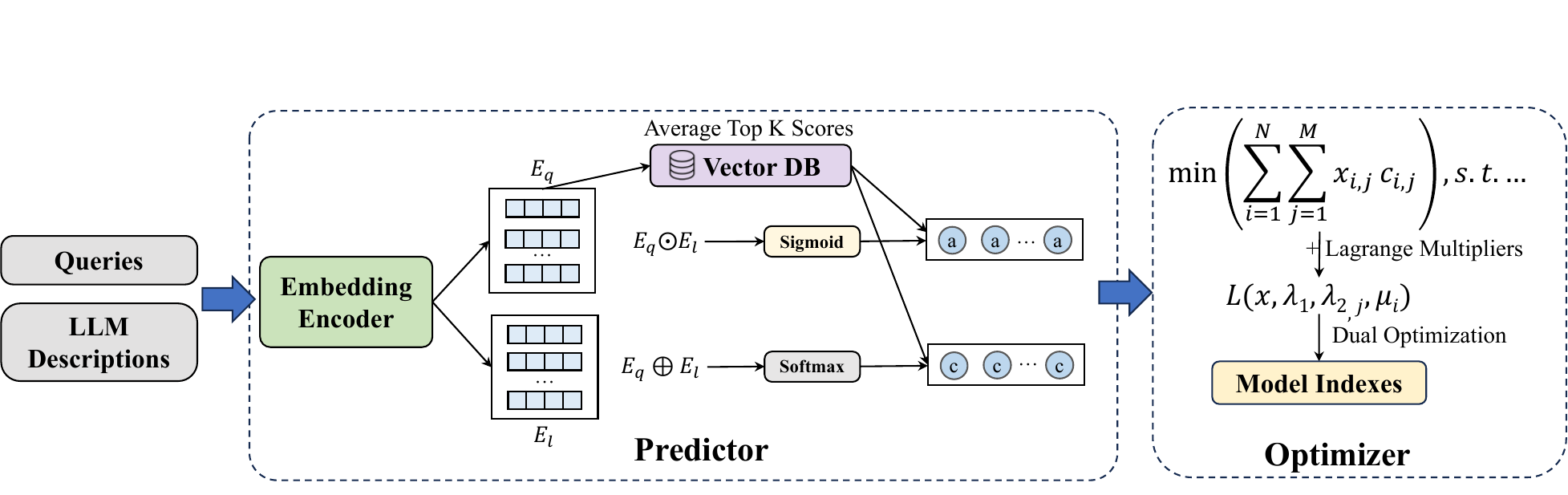}
    \vspace{-8pt}
    
    \caption{Illustration of OmniRouter, including the hybrid predictor and constrained optimizer. }
    \label{fig:model}
\end{figure*}

Inspired by retrieval-augmented generation (RAG) systems \citep{lewis2020retrieval, borgeaud2022improving}, which enhance model outputs by incorporating relevant retrieved information, we design a retrieval-augmented predictor, as illustrated in \autoref{fig:model}, to integrate the generalization capabilities of trained models with the retrieval precision of historical query-model data. 

We employ bert-base-uncased \citep{devlin2018bert} as our embedding encoder due to its representation capabilities and computational efficiency. This encoder processes both queries and LLM descriptions to generate embedding vectors $E_q$ and $E_l$ respectively, which serve as the foundation for both our training-based and retrieval-based prediction components. 
For the training-based component, we design a dual-head architecture built upon the embedding encoder. The first head focuses on model capability prediction, estimating a capability score $a_{i,j}^{pred} \in [0,1]$ for each query $i$ and model $j$ through:
\begin{equation}
a_{i,j}^{pred} = \sigma(\mathbf{W}_1(E_q^i \cdot E_l^j) + \mathbf{b}_1)
\end{equation}
\noindent where $\sigma(\cdot)$ denotes the sigmoid activation function, $\mathbf{W}_1$ and $\mathbf{b}_1$ are learnable parameters. The second head performs sequence length classification by mapping the predicted output token length $l_{i,j}^{pred}$ into discrete buckets $\mathcal{B} = \{B_1, B_2, ..., B_k\}$, where $k = \lceil\frac{l_{max}}{bs}\rceil$, $l_{max}$ is the maximum sequence length, and $bs$ is the bucket size. The probability of length $l_{i,j}^{pred}$ is computed as:
\begin{equation}
l_{i,j}^{pred} = bs \cdot \text{softmax}(\mathbf{W}_2(E_q^i + E_l^j) + \mathbf{b}_2)_i 
\end{equation}

\noindent The bucketing strategy, a technique also employed in previous works \citep{jin2023s3, zheng2024response, fu2024efficient}, transforms the challenging task of precise token-level prediction into a more tractable bucket-level approximation. During training, we optimize these dual objectives using mean squared error (MSE) for capability prediction and cross-entropy loss for length bucket classification.

Concurrently, our retrieval-based component leverages a vector database to identify the top-$k$ similar historical queries based on cosine similarity between query embeddings. For a given query embedding $E_q^i$, let $\mathcal{Q}_k(E_q^i)$ be the set of top-$k$ similar queries retrieved from the database. The retrieval-based capability score $a_{i,j}^{retrieve}$ and output length score $l_{i,j}^{retrieve}$ for query $i$ and model $j$ are computed as:
\begin{small}
\begin{equation}
    l_{i,j}^{retrieve} = \frac{\sum_{q_m \in \mathcal{Q}_k(E_q^i)} \text{sim}(E_q^i, E_{q_m}) \cdot l_{m,j}}{\sum_{q_m \in \mathcal{Q}_k(E_q^i)} \text{sim}(E_q^i, E_{q_m})}
\end{equation}
\end{small}

\begin{small}
\begin{equation}
    a_{i,j}^{retrieve} = \frac{\sum_{q_m \in \mathcal{Q}_k(E_q^i)} \text{sim}(E_q^i, E_{q_m}) \cdot a_{m,j}}{\sum_{q_m \in \mathcal{Q}_k(E_q^i)} \text{sim}(E_q^i, E_{q_m})}
\end{equation}
\end{small}

\noindent where $\text{sim}(E_q^i, E_{q_m})$ denotes the cosine similarity between the current query embedding and the retrieved query embedding, while $l_{m,j}$ and $a_{m,j}$ represent the historically observed output length and capability score for retrieved query $m$ on model $j$, respectively. To obtain the final predictions, we integrate both components through an adaptive fusion mechanism for each query-model pair:
\begin{equation}
    a_{i,j} = \gamma \cdot a_{i,j}^{pred} + (1-\gamma) \cdot a_{i,j}^{retrieve}
\end{equation}

\begin{equation}
    c_{i,j} = \delta \cdot tp_j(l_{i,j}^{pred}) + (1-\delta) \cdot tp_j(l_{i,j}^{retrieve})
\end{equation}

\noindent where $\gamma$ and $\delta \in [0,1]$ are learnable parameters that control the balance between trained and retrieved predictions, and $tp(\cdot)$ is a function that maps the predicted token length to a computational cost estimate based on model $j$'s token prices. This integrated retrieval-augmented architecture ensures our predictor benefits from both generalizable patterns learned during training and specific historical performance data, resulting in more accurate predictions of the uncertain variables needed for the subsequent optimization stage.

\subsection{Constrained Optimizer} \label{sec:optimizer}
At the second stage, we solve the $x_{i,j}$ with the predicted $a_{i,j}$ and $c_{i,j}$. 
Inspired by optimizers \citep{zhang2020solving, bertsekas2014constrained, wang2023multi}, we leverage the Lagrangian dual theory and introduce Lagrangian multipliers to convert the primal problem into its Lagrangian dual problem. By introducing Lagrangian multipliers $\lambda_1,\lambda_{2,j}, \mu_{i}$, we get the Lagrangian relaxation function of the original problem as follows: 
\begin{align}
\min_{x} \quad & \sum_{i=1}^N \sum_{j=1}^M c_{i,j} x_{i,j} \\ \nonumber
\text{s.t.} \quad & \frac{1}{N} \sum_{i=1}^N \sum_{j=1}^M a_{i,j} x_{i,j} \geq \alpha \\ \nonumber
& \sum_{i=1}^N x_{i,j} \leq L_j, \forall j \in [M] & \sum_{j=1}^M x_{i,j} = 1, \forall i \in [N].
\end{align}
we introduce three types of Lagrange multipliers as:
\begin{itemize}
    \item $\lambda_1 \geq 0$ for the quality constraint (inequality)
    \item $\lambda_{2,j} \geq 0$ for each capacity constraint (inequality)
    \item $\mu_i$ for each assignment constraint (equality)
\end{itemize}
and write the Lagrangian function as:
\begin{align}
& L(x,\lambda_1,\lambda_{2,j}, \mu_i) \\ \nonumber
& = \sum_{i=1}^N \sum_{j=1}^M c_{i,j} x_{i,j} + \lambda_1 \left(-\frac{1}{N} \sum_{i=1}^N \sum_{j=1}^M a_{i,j} x_{i,j} + \alpha \right) \\ \nonumber
&\quad + \sum_{j=1}^M \lambda_{2,j} \left(\sum_{i=1}^N x_{i,j} - L_j\right) + \sum_{i=1}^N \mu_i \left(\sum_{j=1}^M x_{i,j} - 1\right)
\end{align}
which can rearranged to group terms with $x_{i,j}$:
\begin{align}
& L(x,\lambda_1,\lambda_{2,j}, \mu_i) \\ \nonumber
& = \sum_{i=1}^N \sum_{j=1}^M x_{i,j}\left(c_{i,j} - \frac{\lambda_1 a_{i,j}}{N} + \lambda_{2,j} + \mu_i\right) \\ \nonumber 
& \quad + \lambda_1\alpha - \sum_{j=1}^M \lambda_{2,j}L_j - \sum_{i=1}^N \mu_i
\end{align}
The KKT optimality conditions for this problem are: 
Stationarity:
\begin{align}
\frac{\partial L}{\partial x_{i,j}} = c_{i,j} - \frac{\lambda_1 a_{i,j}}{N} + \lambda_{2,j} + \mu_i = 0, \quad \forall i,j
\end{align}
Primal Feasibility:
\begin{align}
& \frac{1}{N} \sum_{i=1}^N \sum_{j=1}^M a_{i,j} x_{i,j} \geq \alpha, \quad \sum_{i=1}^N x_{i,j} \leq L_j, \forall j, \quad \sum_{j=1}^M x_{i,j} = 1, \forall i
\end{align}
Dual Feasibility:
\begin{align}
& \lambda_1 \geq 0, \quad \lambda_{2,j} \geq 0, \forall j
\end{align}
Complementary Slackness:
\begin{align}
& \lambda_1 \left(\alpha - \frac{1}{N} \sum_{i=1}^N \sum_{j=1}^M a_{i,j} x_{i,j}\right) = 0, \quad \lambda_{2,j} \left(L_j - \sum_{i=1}^N x_{i,j}\right) = 0, \quad \forall j
\end{align}
Note that for each $i$, we have the constraint: $\sum_{j=1}^M x_{i,j} = 1$. 
This implies that for each query $i$, exactly one LLM $j$ must be selected. From the stationarity condition, for any fixed $i$, comparing two different indices $j$ and $k$:
\begin{align}
& c_{i,j} - \frac{\lambda_1 a_{i,j}}{N} + \lambda_{2,j} + \mu_i = 0 
\end{align}
\begin{align}
c_{i,k} - \frac{\lambda_1 a_{i,k}}{N} + \lambda_{2,k} + \mu_i = 0\end{align}
Subtracting these equations eliminates $\mu_i$:
\begin{align}
(c_{i,j} - \frac{\lambda_1 a_{i,j}}{N} + \lambda_{2,j}) = (c_{i,k} - \frac{\lambda_1 a_{i,k}}{N} + \lambda_{2,k})
\end{align}
This implies that for a given $i$, the optimal solution should choose the $j$ that minimizes $(c_{i,j} - \frac{\lambda_1 a_{i,j}}{N} + \lambda_{2,j})$.

Then the dual function becomes:
\begin{align}
& g(\lambda_1,\lambda_2) \\ \nonumber 
& = \min_{x_{i,j}} \{\sum_{i=1}^N\sum_{j=1}^M x_{i,j}(c_{i,j} - \frac{\lambda_1a_{i,j}}{N} + \lambda_{2,j}) + \lambda_1\alpha - \sum_{j=1}^M \lambda_{2,j}L_j\}
\end{align}
Note that $\mu_i$ has disappeared from the dual function because we've analytically incorporated the equality constraints.
The dual problem can now be written as:
\begin{align}
\max_{\lambda_1,\lambda_2} \quad & g(\lambda_1,\lambda_2), & \text{s.t.} \quad \lambda_1 \geq 0, \quad \lambda_{2,j} \geq 0, \quad \forall j \in [M]
\end{align}
The partial derivatives for the remaining multipliers are:
For $\lambda_1$:
\begin{align}
\frac{\partial L}{\partial \lambda_1} = -\frac{1}{N}\sum_{i=1}^N\sum_{j=1}^M x_{i,j}a_{i,j} + \alpha
\end{align}
For $\lambda_{2,j}$:
\begin{align}
\frac{\partial L}{\partial \lambda_{2,j}} = \sum_{i=1}^N x_{i,j} - L_j, \quad \forall j \in [M]
\end{align}
The gradient ascent update rules are:
\begin{align}
\lambda_1^{t+1} = & \max\left(\lambda_1^t + \alpha_1\left(-\frac{1}{N}\sum_{i=1}^N\sum_{j=1}^M x_{i,j}a_{i,j} + \alpha\right), 0\right) \\
\lambda_{2,j}^{t+1} = & \max\left(\lambda_{2,j}^t + \alpha_2\left(\sum_{i=1}^N x_{i,j} - L_j\right), 0\right), \quad \forall j \in [M]
\end{align}
For fixed multipliers, the optimal assignment for each $i$ is:
\begin{align}
x_{i,j} = \begin{cases}
1 & \text{if } j = j^*_i \\
0 & \text{otherwise}
\end{cases}
\end{align}
where
\begin{align}
j^*_i = \arg\min_{j \in [M]} \left(c_{i,j} - \frac{\lambda_1 a_{i,j}}{N} + \lambda_{2,j}\right)
\end{align}

\begin{itemize}
    \item $\lambda_1$ acts as a penalty for violating the quality constraint. When the average quality is below $\alpha$, $\lambda_1$ increases, encouraging selection of higher-quality options.
    \item $\lambda_{2,j}$ penalizes capacity violations for each model $j$. When a model exceeds its capacity $L_j$, its corresponding multiplier increases, making it less attractive in subsequent iterations.
    \item The equality constraints ($\mu_i$) are handled analytically by direct substitution, which simplifies the dual problem.
    \item The algorithm alternates between updating multipliers and assignments until convergence. 
\end{itemize}
For $\lambda_{2,j}$, its partial derivative shows the workload violation for each model j. If a model's assigned queries exceed its capacity, the gradient is positive, increasing $\lambda_{2,j}$, which makes this overloaded model less attractive in subsequent iterations.

\section{Evaluation} \label{sec:exp}

In this section, we propose the following research questions regarding the performance of OmniRouter and conduct experiments to answer these research questions. 

$\bullet$\; \textbf{RQ1:} What is the routing performance of the OmniRouter and how is compared with existing routing frameworks? 

$\bullet$\; \textbf{RQ2:} Whether the OmniRouter can successfully control budget costs and ensure response accuracy when constraints (i.e., performance constraint $\alpha$ and concurrency constraint $L$) vary? 

$\bullet$\; \textbf{RQ3:} What is the influence of different factors (i.e., modules and parameters in predictors) of OmniRouter on routing performance? 

\subsection{Datasets} \label{sec:dataset}
We collect 2.7k questions sourced from established knowledge and mathematical reasoning datasets, including MMLU \citep{hendrycks2020measuring}, GPQA \citep{rein2023gpqa}, MATH \citep{hendrycks2021measuring}, and GSM8K \citep{cobbe2021training}. We select 10 different models, including 5 relatively weak models, i.e., Qwen2.5 (7B-Instruct, 14B-Instruct, 32B-Instruct) \citep{yang2024qwen2}, Gemma2 (9B-it, 27B-it) \citep{team2024gemma} and 5 relatively strong models, i.e., Qwen2.5-72B-Instruct, GPT 4o-mini, GPT-4o \citep{achiam2023gpt}, Gemini-1.5-flash \citep{team2023gemini}, Claude-3.5-sonnet \footnote{https://claude.ai/}. And we collect the response correctness and token usage of these models to answer these questions using Llama-3.1-70B-Instruct \citep{dubey2024llama} as the evaluation judge \citep{li2024llms}. The Llama-3.1-70B is excluded from the LLM candidate pool to reduce potential biases. The prompt is shown as below. 
\begin{tcolorbox}[
    title=\texttt{Prompt for using LLM as the judge. },
    width=\columnwidth 
]
\begin{flushleft}
\textbf{Prompt:} The ground truth answer is: \{gt\_answer\}. The prediction answer is: {extracted\_answer}. Judge whether the prediction answer is correct or not. You just need to output `True' or `False'. 

\end{flushleft}
\end{tcolorbox}
\label{fig:prompt_in_passkey_LLM}
The statistics of the data we use are in \autoref{tab:sample}. We also set the difficulty for each question, which is determined by the number of models that can answer the question correctly, i.e., Easy: $\{8,9,10\}$ models can answer correctly, Medium: $\{4,5,6,7\}$ models can answer correctly, Hard: $\{0,1,2,3\}$ models can answer correctly. And the proposition is: 78.4\%, 15.2\%, 6.4\%, respectively. 

\begin{table}[htbp]
\centering
\caption{Statistics of the data distribution in our dataset. } \label{tab:sample}
\vspace{-8pt}
\begin{tabular}{ccc}
\toprule
Data Source  & \# Samples & Ratio \\ \midrule
MMLU         & 1000       & 37.06\% \\
GPQA-Diamond & 198        & 7.33\%  \\
Math-500     & 500        & 18.53\% \\
GSM8K        & 1000        & 37.06\% \\ \bottomrule
\end{tabular}
\end{table}

For calculating costs for LLMs to answer queries, we refer to litellm\footnote{https://docs.litellm.ai/docs/} for calculating money cost, the money price regarding the model we use are shown in \autoref{tab:model_pricing}. 

\begin{table}[tb!]
    \centering
    \caption{Money Cost Map of LLMs}
    \label{tab:model_pricing}
    \vspace{-8pt}

    \adjustbox{max width=1.0\linewidth}{\begin{tabular}{lcc}
        \toprule
        \textbf{Model Name} & \textbf{1M Input Tokens (\$)} & \textbf{1M Output Tokens (\$)} \\
        \midrule
        Qwen-2.5-7B-Instruct \citep{yang2024qwen2} & 0.267 & 0.267 \\
        Qwen-2.5-14B-Instruct \citep{yang2024qwen2} & 0.534 & 0.534 \\
        Qwen-2.5-32B-Instruct \citep{yang2024qwen2} & 1.22 & 1.22 \\
        Qwen-2.5-72B-Instruct \citep{yang2024qwen2} & 2.745 & 2.745 \\
        Gemma-2-9B-it \citep{team2024gemma} & 0.343 & 0.343 \\
        Gemma-2-27B-it \citep{team2024gemma} & 1.03 & 1.03 \\
        gpt-4o-mini \citep{achiam2023gpt} & 0.15 & 0.6 \\
        gpt-4o \citep{achiam2023gpt} & 2.5 & 10 \\
        gemini-1.5-flash \citep{team2023gemini} & 0.075 & 0.3 \\
        claude-3-5-sonnet\footnote{https://www.anthropic.com/news/claude-3-5-sonnet} & 3 & 15 \\
        \bottomrule
    \end{tabular}}
\end{table}

\subsection{Setup} \label{sec:setup}
We conduct our experiments on an Ubuntu 22.04 machine equipped with 8 RTX A5000 GPUs to serve multiple LLMs using Ollama \footnote{https://ollama.com/}. During the implementation, we set the maximum output length to 1024 tokens to prevent the influence of extremely long output sequences, set top $k$ as 16 during retrieval, and $\gamma$ and $\delta$ as 0.5 during aggregation.  
To calculate money costs, we refer to Litellm's costmap for price reference. The detailed price map can be found in Appendix B. 
Our experiments are conducted on the continuous-batching setting \citep{kwon2023efficient}. The serving system processes requests by dynamically forming batches based on the incoming query traffic, providing a more realistic evaluation environment for routing frameworks. In our simulation, we model real-world traffic patterns by randomly adding $n \in \{1,2,3,4\}$ queries to the queue every 0.1 seconds and performing routing decisions at 1-second intervals. 
If not specifically mentioned, we set the two constraints $\alpha=0.75$ and the concurrent workload constraint $L=4$, which are applied uniformly across all LLMs. Controllability analysis of different values of $\alpha$ and $L$ is presented in Section \ref{sec:control}.
We evaluate OmniRouter against the following routing baselines: 

\textbf{Cost-oriented:} We adapt S3 \citep{jin2023s3} and PO \citep{zheng2024response}, which are originally designed for latency optimization as cost-oriented baselines. Specifically, S3 employs DistilBERT \citep{sanh2019distilbert} and PO employs Vicuna-7B \citep{chiang2023vicuna} as output token length predictors, respectively.

\textbf{Performance-oriented:} For performance-focused routing, we implement EmbedLLM \citep{zhuang2024embedllm} and RouterDC \citep{chen2024routerdc}. EmbedLLM leverages an encoder-decoder architecture to embed LLM representations, while RouterDC employs contrastive learning to model query-LLM relationships, both aiming to maximize response quality regardless of computational cost.

\textbf{Cost-performance Coordinated:} For baselines which balance both cost and performance objectives, we employ Hybrid-LLM \citep{hybridllm2024}, which constructs a probabilistic router, and CARROT \citep{carrot2025}, which implements dual predictors using Llama3-8B \citep{dubey2024llama} as specified in its original implementation. 

\subsection{RQ1: Routing Performance}
\begin{table*}[tb!]
\centering
\caption{Routing performance comparison, where we use accuracy, i.e., whether the routed LLM successfully answers the assigned query, and money cost of calling respective LLMs. } \label{tab:overall}
\vspace{-8pt}
\adjustbox{max width=1.0\linewidth}{
\begin{tabular}{lcccccccc}
\toprule
Metric & S3 & PO & EmbedLLM & RouterDC & Hybrid-LLM & CARROT & \textbf{OmniRouter} \\ 
\midrule
Accuracy ($\uparrow$) & 69.45\% & 68.71\% & 72.96\% & 73.89\% & 71.48\% & 72.41\% & \textbf{75.19\%} \\
\$ Cost ($\downarrow$) & 0.0585 & 0.0610 & 0.0896 & 0.0874 & 0.0722 & 0.0680 & \textbf{0.0515} \\
\bottomrule
\end{tabular}
}
\end{table*}
The comparison of overall routing performance can be observed from \autoref{tab:overall}. OmniRouter demonstrates substantial advantages over all baselines across both metrics. Given the theoretical performance bounds (lower bound is 57.41\% when the worst LLM for each query is selected and upper bound is 90\% when the best LLM for each query is selected), the improvement achieved by OmniRouter (1.3\% to 6.48\% absolute improvement) is particularly significant, which represents a meaningful advancement within this constrained optimization space.
When compared to cost-oriented baselines (S3 and PO), OmniRouter achieves significantly higher success rates (5.74\% and 6.48\% improvements respectively) while maintaining comparable or even lower costs. Against performance-oriented approaches such as EmbedLLM and RouterDC, our method achieves both higher success rates and substantially lower costs (approximately 41\% cost reduction compared to RouterDC). Even when compared to cost-performance coordinated methods (Hybrid-LLM and CARROT), OmniRouter demonstrates superior performance in both dimensions.

\begin{figure}[tb!]
    \centering
    \includegraphics[width=0.9\linewidth]{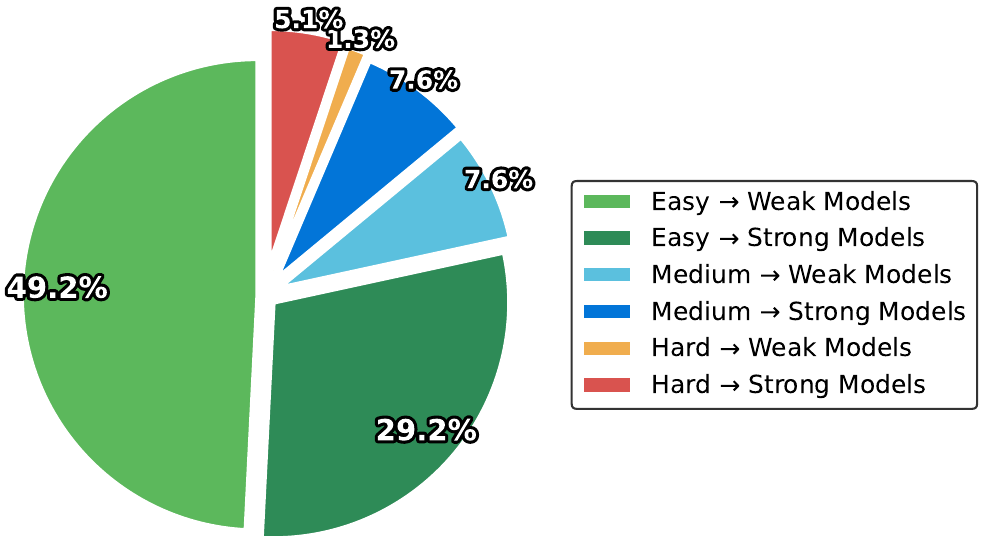}
    \vspace{-8pt}
    \captionof{figure}{Distribution of OmniRouter's query routing decisions across difficulty levels and model capabilities.}
    \label{fig:query_distribution}
\end{figure}

\begin{table}[tb!]
    \centering
    \captionof{table}{Performance of predictors on model capability and output token length prediction.}
    \vspace{-8pt}
    \label{tab:predictor}
    \adjustbox{max width=1.0\linewidth}{
    \begin{tabular}{lccc}
    \toprule
    \multirow{2}{*}{Method} & \multirow{2}{*}{Capability Acc.} & \multicolumn{2}{c}{Length Bucket Acc.} \\ \cmidrule(l){3-4} 
                            &                                  & Exact Match        & $\pm1$       \\ \midrule
    S3                      & --                                & 0.333              & 0.656          \\
    PO                      & --                                & 0.325              & 0.683          \\
    EmbedLLM                & 0.732                             & --                 & --             \\
    RouterDC                & 0.761                             & --                 & --             \\
    Hybrid-LLM              & 0.708                             & --                 & --             \\
    CARROT                  & 0.745                             & 0.312              & 0.672          \\
    \midrule
    \textbf{OmniRouter}     & \textbf{0.813}                    & \textbf{0.452}     & \textbf{0.806} \\
    \bottomrule
    \end{tabular}
    }
\end{table}

\textbf{Performance of Predictors.}
We also compare the effectiveness of OmniRouter's predictor with the baselines we have mentioned. \autoref{tab:predictor} demonstrates OmniRouter's significant advantages in prediction accuracy. Our approach achieves 81.3\% capability prediction accuracy, outperforming the strongest baseline (RouterDC) by 5.2\%. For length prediction, OmniRouter delivers 45.2\% exact match accuracy and 80.6\% relaxed ($\pm 1$) accuracy, surpassing all alternatives by over 13.0\%. 

\textbf{Routing Performance across Query Difficulty Levels.}
To assess the routing performance of OmniRouter deeper, we analyze its allocation patterns across different difficulty categories. \autoref{fig:query_distribution} reveals OmniRouter's sophisticated routing strategy across query difficulty levels. For easy queries (78.4\% of workload), the system successfully routes 49.2\% of them to weak models. Medium-difficulty queries receive balanced treatment with equal distribution between model types (7.6\% each). For hard queries, OmniRouter shows a strong preference for routing to strong models by a 4:1 ratio (5.1\% vs 1.3\%), demonstrating the capability of OmniRouter to match appropriate models. 
\subsection{RQ2: Controllability Analysis} \label{sec:control}
We investigate how different routing approaches respond to varying operational constraints, i.e., performance and concurrent constraints. 
\begin{figure}[tb!]
    \centering
    \begin{subfigure}{0.49\linewidth}
        \centering
        \includegraphics[width=\linewidth]{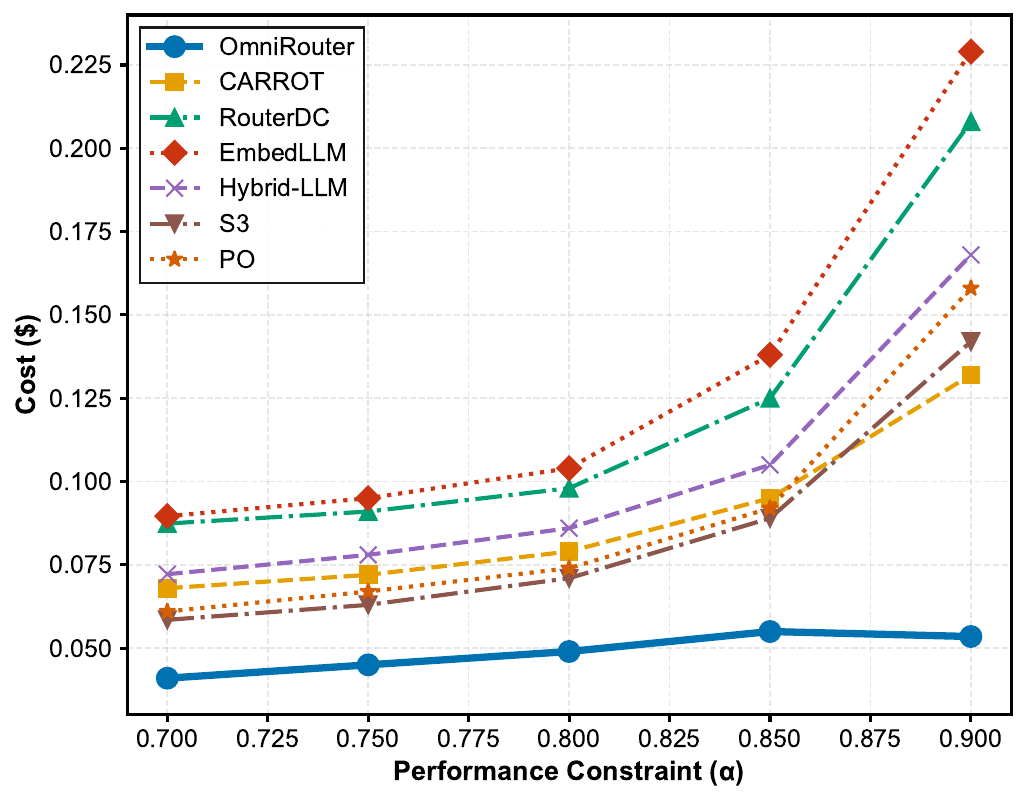}
        \caption{Cost vs. Performance Constraint}
        \label{fig:perf_cost}
    \end{subfigure}
    \hfill
    \begin{subfigure}{0.49\linewidth}
        \centering
        \includegraphics[width=\linewidth]{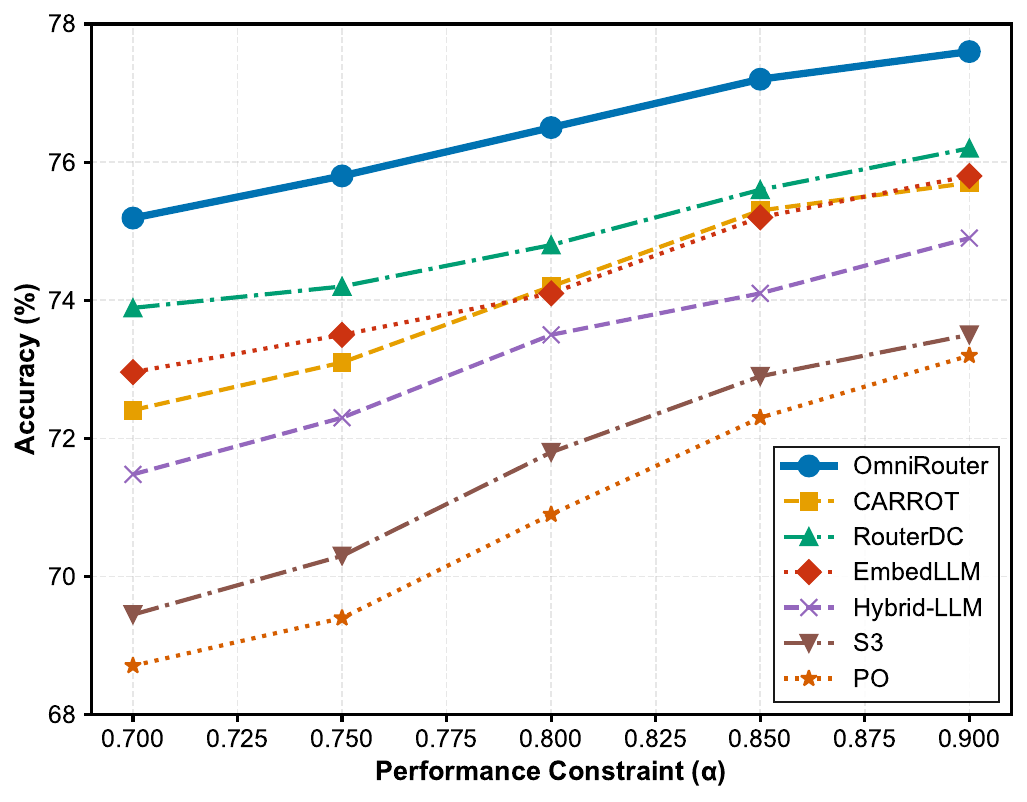}
        \caption{Accuracy vs. Performance Constraint}
        \label{fig:perf_acc}
    \end{subfigure}
    
    \caption{Impact of performance constraint ($\alpha$) on cost efficiency and routing accuracy. As performance requirements increase, greedy methods exhibit unbounded cost growth while OmniRouter's constraint optimization maintains controlled scaling.}
    \label{fig:perf_constraint}
\end{figure}
\autoref{fig:perf_constraint} reveals a critical limitation of existing greedy routing approaches: their inability to bound costs under stringent performance requirements. As the performance constraint ($\alpha$) increases from 0.70 to 0.90, baseline methods exhibit exponential cost growth: at $\alpha=0.90$, the cost of EmbedLLM reaches \$0.229, more than four times OmniRouter's cost of \$0.054, creating a substantial cost gap ($\Delta = 0.175$). This cost explosion occurs because greedy routing strategies inherently prioritize immediate performance gains when performance thresholds rise, defaulting to routing increasingly more queries to the most powerful (and expensive) models regardless of actual query complexity.
In contrast, OmniRouter demonstrates remarkably controlled cost scaling, with only a 48\% increase over the same range and even a slight cost reduction at $\alpha > 0.85$. 

\begin{figure}[tb!]
    \centering
    \begin{subfigure}{0.49\linewidth}
        \centering
        \includegraphics[width=\linewidth]{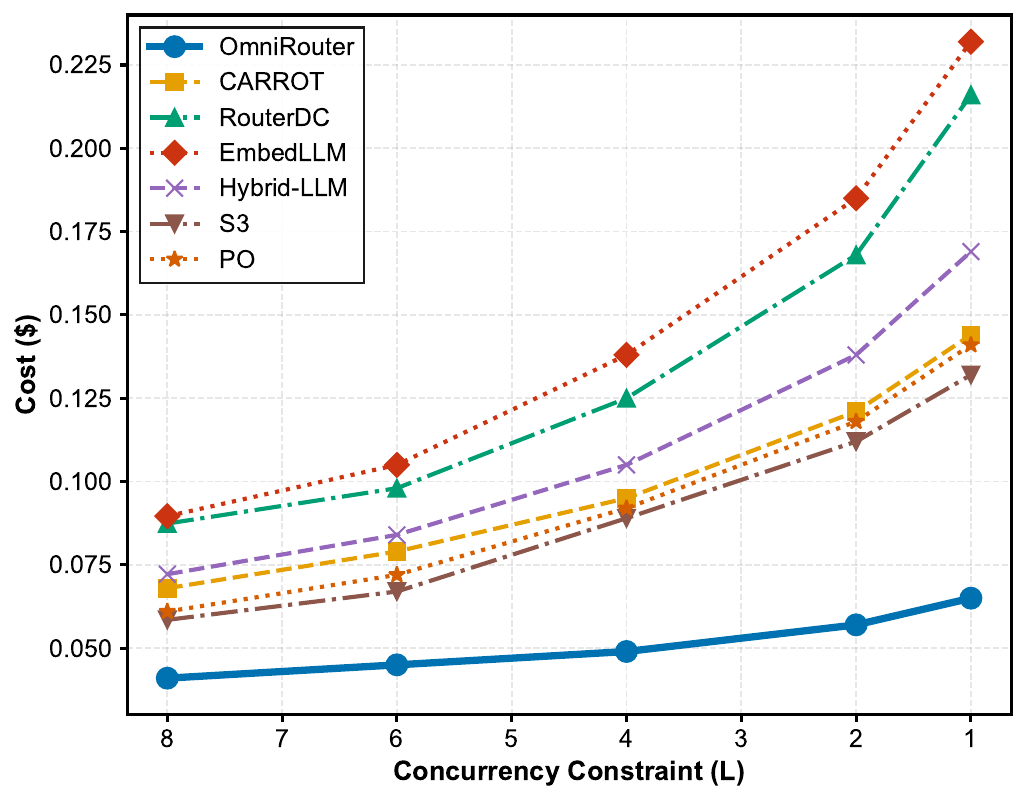}
        \caption{Cost vs. Concurrency Constraint}
        \label{fig:conc_cost}
    \end{subfigure}
    \hfill
    \begin{subfigure}{0.49\linewidth}
        \centering
        \includegraphics[width=\linewidth]{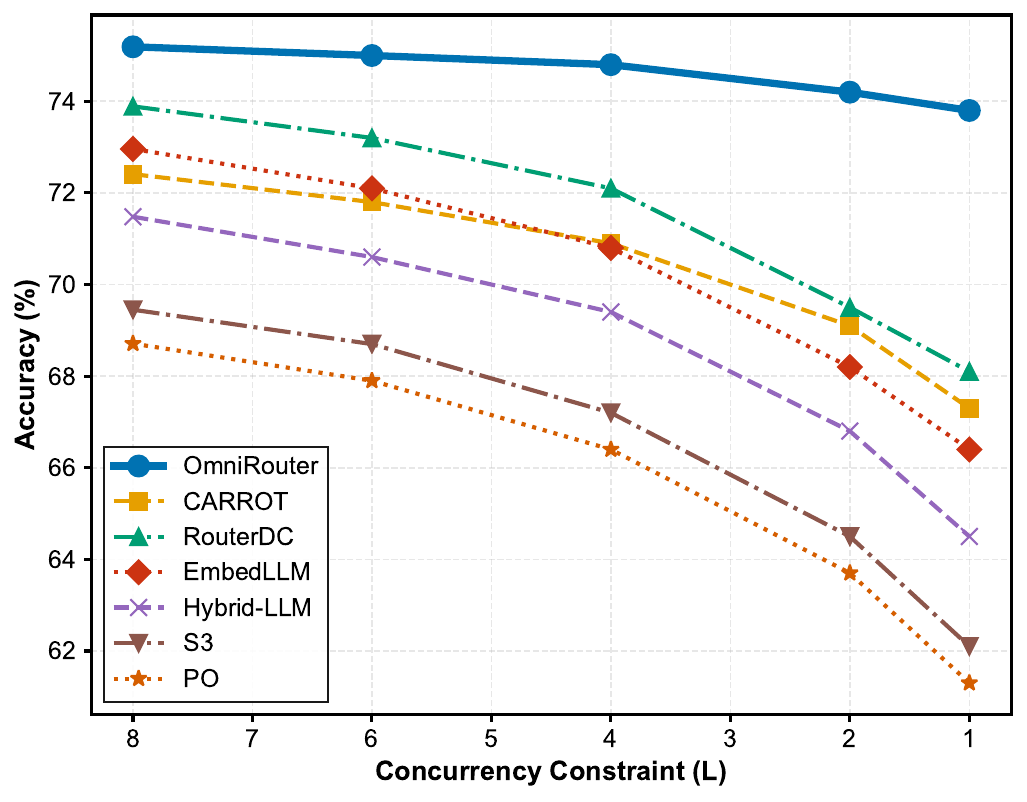}
        \caption{Accuracy vs. Concurrency Constraint}
        \label{fig:conc_acc}
    \end{subfigure}
    \caption{Impact of concurrency constraint ($L$) on cost efficiency and routing accuracy. As available parallelism decreases, greedy methods struggle to make effective compromises, while OmniRouter's constraint optimization maintains balanced allocations.}
    \label{fig:conc_constraint}
\end{figure}

Similarly, in \autoref{fig:conc_constraint}, it demonstrates system behavior under varying concurrency constraints, revealing another key weakness of greedy approaches. As available concurrency decreases from 8 to 1, baseline methods show substantial cost increases (EmbedLLM: 159\%, RouterDC: 147\%) and significant performance degradation. When concurrency is highly limited ($L=1$), the performance gap between OmniRouter and the weakest baseline (S3) becomes dramatic ($\Delta = 11.7\%$). This occurs because greedy strategies struggle to make effective compromises when resource constraints tighten. By constrast, OmniRouter substantially performs better, maintaining 73.8\% accuracy even at $L=1$ while experiencing only moderate cost increases.

\subsection{RQ3: Ablation Studies} \label{sec:ablation}
\textbf{Effects of Removing Modules in the Predictor.}
To evaluate the contribution of each component in our retrieval-augmented predictor, we conduct ablation studies by removing key modules from the full model. Results are presented in \autoref{tab:ablation_predictor}. 

\begin{table}[tb!]
\centering
\captionof{table}{Effects of removing different modules in the predictor on routing performance.}
\vspace{-8pt}

\label{tab:ablation_predictor}
\adjustbox{max width=1.0\linewidth}{
\begin{tabular}{lcccc}
\toprule
Predictor & Capability Acc ($\uparrow$) & Bucket Acc ($\uparrow$) & Acc ($\uparrow$) & \$ Cost ($\downarrow$) \\
\midrule
Full & \textbf{0.813} & \textbf{0.806} & \textbf{75.19\%} & \textbf{0.0515} \\
w/o Retrieval & 0.751 & 0.724 & 71.85\% & 0.0698 \\
w/o Training & 0.728 & 0.682 & 73.33\% & 0.0642 \\
\bottomrule
\end{tabular}
}
\end{table}
Removing the retrieval component leads to a substantial drop in both capability accuracy (7.6\% decrease from 0.813 to 0.751) and length prediction accuracy (10.2\% decrease from 0.806 to 0.724). This degradation directly impacts routing effectiveness, resulting in lower success rate (4.4\% decrease) and significantly higher operational costs (35.5\% increase from \$0.0515 to \$0.0698). These findings highlight that historical query information provides critical contextual signals that enhance prediction. On the other hand, 
after removing the training-based component, although performance accuracy sees a larger decline (10.5\% decrease to 0.728) compared to the no-retrieval variant, the success rate decreases less dramatically (only 2.5\% reduction). This suggests that the retrieval mechanism alone sometimes can maintain reasonable routing decisions, which further validate our retrieval-augmented choice to enhance predictors. 

\textbf{Effects of Key Parameters within OmniRouter's Predictors.}
We investigate the sensitivity of OmniRouter's routing effectiveness to key design parameters within its prediction components. As shown in \autoref{tab:n_bucket}, the number of buckets used for discretizing output length predictions significantly impacts routing performance. A smaller number of buckets (10) yields substantially higher prediction accuracy (45.2\% exact match, 80.6\% $\pm1$ accuracy), which directly leads to higher routing performance (75.19\% success rate) and lower operational costs (\$0.0515). This finding suggests that coarser-grained discretization may be sufficient for effective routing decisions, while excessive granularity can introduce noise.
Regarding the retrieval component, \autoref{tab:n_k} demonstrates how varying $K$ (the number of historical samples) affects prediction accuracy and routing performance. The system achieves optimal performance at $K=16$, with the highest success rate (75.19\%) and lowest cost (\$0.0515). We observe that prediction accuracy remains relatively stable across moderate $K$ values (8-32), with only slight degradation at the extremes. Too few samples ($K=4$) provide insufficient historical context, while too many ($K=64$) likely introduce noise from less relevant queries, both resulting in suboptimal routing decisions. This indicates the retrieval-augmentation requires careful calibration to balance between relevant historical information and potentially misleading outliers.


\begin{table}[tb!]
\centering
\centering
\captionof{table}{Effects of number of buckets used in predictors on routing performance.} \label{tab:n_bucket}
\vspace{-8pt}

\adjustbox{max width=1.0\linewidth}{
\begin{tabular}{ccccc}
\toprule
\multirow{2}{*}{\# Bucket} & \multicolumn{2}{c}{Bucket Acc ($\uparrow$)} & \multirow{2}{*}{Capability Acc. ($\uparrow$)} & \multirow{2}{*}{\$ Cost ($\downarrow$)} \\ \cmidrule(lr){2-3}
                           & Exact Match               & $\pm1$              &                                 &                                        \\ \midrule
10                         & \textbf{0.452}            & \textbf{0.806}      & \textbf{75.19\%}                         & \textbf{0.0515}                        \\
20                         & 0.269                     & 0.526               & 73.22\%                & 0.0725                                 \\
50                         & 0.162                     & 0.299               & 73.22\%                         & 0.834                                 \\
100                        & 0.124                     & 0.189               & 70.74\%                         & 0.0962                                 \\ \bottomrule
\end{tabular}
}
\end{table}

\begin{table}[h]
\centering
\captionof{table}{Effects of variations of K in historical samples on routing performance.} \label{tab:n_k}
\vspace{-8pt}

\adjustbox{max width=1.0\linewidth}{
\begin{tabular}{cccccc}
\toprule
\multirow{2}{*}{K} & \multirow{2}{*}{Capability Acc. ($\uparrow$)} & \multicolumn{2}{c}{Length Acc ($\uparrow$)} & \multirow{2}{*}{SR ($\uparrow$)} & \multirow{2}{*}{\$ Cost ($\downarrow$)} \\ \cmidrule(lr){3-4}
                           &                                         & Exact Match      & $\pm1$       &                                 &                                        \\ \midrule
4                          & 0.773                                   & 0.430            & 0.784        & 74.07\%                         & 0.0614                                 \\
8                          & 0.785                          & 0.443            & 0.786        & 74.81\%                         & 0.0565                                 \\
16                         & \textbf{0.806}                                   & \textbf{0.452}   & \textbf{0.806}& \textbf{75.19\%}                & \textbf{0.0515}                        \\
32                         & 0.785                                   & 0.440            & 0.776        & 74.81\%                         & 0.0565                                 \\
64                         & 0.780                                   & 0.422            & 0.764        & 73.70\%                         & 0.0684                                 \\ \bottomrule
\end{tabular}
}
\end{table}

\section{Theoretical Proof}~\label{app:proof}

\textbf{Proposition 1.}
\emph{Given the Lagrangian dual problem with multipliers $\lambda_1 \geq 0$ and $\lambda_{2,j} \geq 0$ for $j \in [M]$, the optimal assignment $x^*_{i,j}$ for the constrained optimization problem is uniquely determined by:}
\begin{equation}
x^*_{i,j} = \begin{cases} 
1 & \text{if } j = \arg\min_{j' \in [M]} \left(c_{i,j'} - \frac{\lambda_1^* a_{i,j'}}{N} + \lambda_{2,j'}^*\right) \\
0 & \text{otherwise}
\end{cases}
\end{equation}

\begin{proof}
We start with the original optimization problem:
\begin{align}
\min_{x} \quad & \sum_{i=1}^N \sum_{j=1}^M c_{i,j} x_{i,j} \\ \nonumber
\text{s.t.} \quad & \frac{1}{N}\sum_{i=1}^N \sum_{j=1}^M a_{i,j} x_{i,j} \geq \alpha \quad \sum_{i=1}^N x_{i,j} \leq L_j, \quad \forall j \in [M] \\ \nonumber
& \sum_{j=1}^M x_{i,j} = 1, \quad \forall i \in [N] \quad x_{i,j} \in \{0,1\}, \quad \forall i,j
\end{align}

The Lagrangian function is:
\begin{align}
& L(x,\lambda_1,\lambda_2,\mu) \\ \nonumber
&= \sum_{i=1}^N \sum_{j=1}^M c_{i,j} x_{i,j} + \lambda_1 \left( \alpha - \frac{1}{N}\sum_{i=1}^N \sum_{j=1}^M a_{i,j} x_{i,j} \right) \\ \nonumber
&\quad + \sum_{j=1}^M \lambda_{2,j} \left( \sum_{i=1}^N x_{i,j} - L_j \right) + \sum_{i=1}^N \mu_i \left( \sum_{j=1}^M x_{i,j} - 1 \right)
\end{align}

From the KKT stationarity condition:
\begin{equation}
\frac{\partial L}{\partial x_{i,j}} = c_{i,j} - \frac{\lambda_1 a_{i,j}}{N} + \lambda_{2,j} + \mu_i = 0
\end{equation}

where $\mu_i$ is the Lagrangian multiplier for the constraint $\sum_{j=1}^M x_{i,j} = 1$.

Rearranging the stationarity condition:
\begin{equation}
\mu_i = -\left(c_{i,j} - \frac{\lambda_1 a_{i,j}}{N} + \lambda_{2,j}\right)
\end{equation}

Since $\sum_{j=1}^M x_{i,j} = 1$ and $x_{i,j} \in \{0,1\}$, exactly one $x_{i,j}$ equals 1 for each query $i$. For the optimal solution, we choose $j^*$ that minimizes the expression in the stationarity condition:
\begin{equation}
j^* = \arg\min_{j \in [M]} \left(c_{i,j} - \frac{\lambda_1 a_{i,j}}{N} + \lambda_{2,j}\right)
\end{equation}

Therefore, $x_{i,j^*} = 1$ and $x_{i,j} = 0$ for $j \neq j^*$, which completes the proof.
\end{proof}

\textbf{Proposition 2.}
\emph{Given the optimal dual variable $\lambda_1^*$ obtained from the gradient ascent update, if $\lambda_1^* > 0$, then the quality constraint is active and satisfied with equality:}
\begin{equation}
\frac{1}{N}\sum_{i=1}^N \sum_{j=1}^M a_{i,j} x_{i,j}^* = \alpha
\end{equation}

\begin{proof}
From the complementary slackness condition in the KKT conditions:
\begin{equation}
\lambda_1^* \left( \alpha - \frac{1}{N}\sum_{i=1}^N \sum_{j=1}^M a_{i,j} x_{i,j}^* \right) = 0
\end{equation}

This equation states that either:
\begin{itemize}
    \item[(i)] $\lambda_1^* = 0$, or
    \item[(ii)] $\alpha - \frac{1}{N}\sum_{i=1}^N \sum_{j=1}^M a_{i,j} x_{i,j}^* = 0$
\end{itemize}

Given that $\lambda_1^* > 0$, condition (i) is false, thus condition (ii) must hold:
\begin{equation}
\alpha - \frac{1}{N}\sum_{i=1}^N \sum_{j=1}^M a_{i,j} x_{i,j}^* = 0
\end{equation}

Rearranging yields:
\begin{equation}
\frac{1}{N}\sum_{i=1}^N \sum_{j=1}^M a_{i,j} x_{i,j}^* = \alpha
\end{equation}

This demonstrates that when $\lambda_1^* > 0$, the quality constraint is satisfied with equality. The dual variable $\lambda_1^*$ acts as a "price" for quality: the optimizer finds the solution that exactly meets the quality threshold $\alpha$ while minimizing cost.
\end{proof}

\section{Related Work}
\textbf{LLM Routing.} To optimize performance and inference cost in multi-LLM serving systems, researchers have developed increasingly sophisticated routing approaches. 
RouteLLM \citep{ong2024routellm} and CARROT \citep{carrot2025} propose training small LLMs as routers to balance cost and performance. EmbedLLM \citep{zhuang2024embedllm} introduces a specialized encoder-decoder network for embedding LLM representations. HybridLLM \citep{hybridllm2024} addresses data imbalance issues by proposing a probabilistic router to better represent different model capabilities across query types. C2MAB-V \citep{dai2024cost} treats routing as a contextual multi-armed bandit problem, enabling exploration-exploitation trade-offs in model selection. GraphRouter \citep{feng2024graphrouter} leverages graph learning to jointly modeling the query-model, query-query, and model-model relationship for building routers. While these approaches have made substantial progress, they predominantly operate as greedy decision-makers, optimizing for individual queries without considering global system constraints or the implications for subsequent requests. DeepSieve~\citep{guo2025deepsieve} utilizes LLM-as-a-Knowledge-Router to enable effective reasoning across multiple heterogeneous knowledge sources, enhancing retrieval-augmented generation (RAG) systems by dynamically routing sub-queries to the most appropriate source.

\textbf{LLM Generation Length Prediction. }
Predicting LLM generation length is crucial for optimizing computational resources. Early attempts like Magnus \citep{cheng2024enabling} employed random forest algorithms but achieved limited accuracy. Subsequent research has explored two main directions of prediction models: encoder-only models for classification (DynamoLLM \citep{stojkovic2024dynamollm}, S3 \citep{jin2023s3}, TerriInfer \citep{hu2024inference}, SSJF \citep{aiops2024qiu}, and $\mu$3 \citep{qiu2024power}) and decoder-only models for generative prediction like Perception-only (PO) \citep{zheng2024response}. LTR \citep{fu2024efficient} reformulated this as a ranking problem and utilized listwise ranking for predictor training. 

\section{Conclusion}
In this paper, we introduce OmniRouter, a routing framework that fundamentally reframes LLM routing as a constrained optimization problem rather than a series of greedy decisions. We develop a two-stage approach with a hybrid predictor that accurately estimates model capabilities and costs, and a constrained optimizer that minimizes operational expenses while satisfying performance requirements.
Our experiments with the \dataset dataset demonstrate that OmniRouter achieves up to 6.30\% higher response accuracy while reducing costs by at least 10.15\% compared to existing methods. OmniRouter maintains remarkable stability under varying constraints, precisely where greedy approaches fail. These results confirm OmniRouter's effectiveness for more realistic environments requiring both performance guarantees and cost control.


\bibliographystyle{acm}
\bibliography{ref}

@article{jin2025two,
  title={Two heads are better than one: Test-time scaling of multi-agent collaborative reasoning},
  author={Jin, Can and Peng, Hongwu and Zhang, Qixin and Tang, Yujin and Metaxas, Dimitris N and Che, Tong},
  journal={arXiv preprint arXiv:2504.09772},
  year={2025}
}

@article{panda2025adaptive,
  title={Adaptive llm routing under budget constraints},
  author={Panda, Pranoy and Magazine, Raghav and Devaguptapu, Chaitanya and Takemori, Sho and Sharma, Vishal},
  journal={arXiv preprint arXiv:2508.21141},
  year={2025}
}

@article{yang2024qwen2,
  title={Qwen2. 5 technical report},
  author={Yang, An and Yang, Baosong and Zhang, Beichen and Hui, Binyuan and Zheng, Bo and Yu, Bowen and Li, Chengyuan and Liu, Dayiheng and Huang, Fei and Wei, Haoran and others},
  journal={arXiv preprint arXiv:2412.15115},
  year={2024}
}

@article{chiang2023vicuna,
  title={Vicuna: An open-source chatbot impressing gpt-4 with 90\%* chatgpt quality},
  author={Chiang, Wei-Lin and Li, Zhuohan and Lin, Ziqing and Sheng, Ying and Wu, Zhanghao and Zhang, Hao and Zheng, Lianmin and Zhuang, Siyuan and Zhuang, Yonghao and Gonzalez, Joseph E and others},
  journal={See https://vicuna. lmsys. org (accessed 14 April 2023)},
  volume={2},
  number={3},
  pages={6},
  year={2023}
}

@article{dai2024cost,
  title={Cost-effective online multi-llm selection with versatile reward models},
  author={Dai, Xiangxiang and Li, Jin and Liu, Xutong and Yu, Anqi and Lui, John},
  journal={arXiv preprint arXiv:2405.16587},
  year={2024}
}

@article{sanh2019distilbert,
  title={DistilBERT, a distilled version of BERT: smaller, faster, cheaper and lighter},
  author={Sanh, Victor and Debut, Lysandre and Chaumond, Julien and Wolf, Thomas},
  journal={arXiv preprint arXiv:1910.01108},
  year={2019}
}

@article{team2024gemma,
  title={Gemma 2: Improving open language models at a practical size},
  author={Team, Gemma and Riviere, Morgane and Pathak, Shreya and Sessa, Pier Giuseppe and Hardin, Cassidy and Bhupatiraju, Surya and Hussenot, L{\'e}onard and Mesnard, Thomas and Shahriari, Bobak and Ram{\'e}, Alexandre and others},
  journal={arXiv preprint arXiv:2408.00118},
  year={2024}
}

@inproceedings{zhang2023s,
  title={S-LoRA: Serving Thousands of Concurrent LoRA Adapters},
  author={Zhang, Ying and Shi, Xuangui and Feng, Leyang and Wang, Min and Yu, Yang and Yu, Xiwei and Shen, Hailin and Chen, Zhifang and Mucci, Phillip and Kudlur, Manjunath and others},
  booktitle={Advances in Neural Information Processing Systems},
  volume={36},
  year={2023}
}

@inproceedings{feng2023towards,
  title={Towards Understanding and Mitigating the Training Data Quality of Large Language Models},
  author={Feng, Xianjun and Wu, Mengjie and Feng, Yunlong and Yin, Xiaomeng and Wang, Shuiqiao and Zeng, Ying and Zeng, Xiangzhuo and Wang, Zheyuan and Qin, Ruiyi and Hu, Guocheng and others},
  booktitle={Advances in Neural Information Processing Systems},
  volume={36},
  year={2023}
}

@inproceedings{qin2023towards,
  title={Towards Robust LLM-based Decision-Making: A Calibration and Planning Approach},
  author={Qin, Xingchen and Yang, Yuxi and Li, Xuming and Dong, Sicen and Huang, Shujian and Ji, Heng and Li, Lei},
  booktitle={International Conference on Learning Representations},
  year={2023}
}

@inproceedings{yu2022gilbo,
  title={GILBO: One Metric to Measure Them All},
  author={Yu, Peilin and Trombetta, Patrick and Hassani, Ahmad and Bulitko, Vadim and White, Martha},
  booktitle={Advances in Neural Information Processing Systems},
  volume={35},
  pages={10285--10297},
  year={2022}
}

@inproceedings{borgeaud2022improving,
  title={Improving language models by retrieving from trillions of tokens},
  author={Borgeaud, Sebastian and Mensch, Arthur and Hoffmann, Jordan and Cai, Trevor and Rutherford, Eliza and Millican, Katie and Van Den Driessche, George Bornea and Lespiau, Jean-Baptiste and Damoc, Bogdan and Clark, Aidan and others},
  booktitle={International Conference on Machine Learning},
  pages={2206--2240},
  year={2022},
  organization={PMLR}
}

@article{ong2024routellm,
  title={RouteLLM: Learning to Route LLMs with Preference Data},
  author={Isaac Ong and Amjad Almahairi and Vincent Wu and Wei-Lin Chiang and Tianhao Wu and Joseph E. Gonzalez and M Waleed Kadous and Ion Stoica},
  journal={arXiv preprint arXiv:2406.18665},
  year={2024}
}

@article{hybridllm2024,
  title={Hybrid LLM: Cost-Efficient and Quality-Aware Query Routing},
  author={Anonymous},
  journal={arXiv preprint arXiv:2404.14618},
  year={2024}
}

@article{carrot2025,
  title={CARROT: A Cost Aware Rate Optimal Router},
  author={Seamus Somerstep and Felipe Maia Polo and Allysson Flavio Melo de Oliveira and Prattyush Mangal and Mírian Silva and Onkar Bhardwaj and Mikhail Yurochkin and Subha Maity},
  journal={arXiv preprint arXiv:2502.03261},
  year={2025}
}

@article{zhang2025capability,
  title={Capability Instruction Tuning: A New Paradigm for Dynamic LLM Routing},
  author={Yi-Kai Zhang and De-Chuan Zhan and Han-Jia Ye},
  journal={arXiv preprint arXiv:2502.17282},
  year={2025}
}

@article{zhuang2024embedllm,
  title={EmbedLLM: Learning Compact Representations of Large Language Models},
  author={Zhuang, Richard and Wu, Tianhao and Wen, Zhaojin and Li, Andrew and Jiao, Jiantao and Ramchandran, Kannan},
  journal={arXiv preprint arXiv:2410.02223},
  year={2024}
}

@article{parkar2024selectllm,
  title={SelectLLM: Can LLMs Select Important Instructions to Annotate?},
  author={Parkar, Ritik Sachin and Kim, Jaehyung and Park, Jong Inn and Kang, Dongyeop},
  journal={arXiv preprint arXiv:2401.16553},
  year={2024}
}

@article{feng2024graphrouter,
  title={Graphrouter: A graph-based router for llm selections},
  author={Feng, Tao and Shen, Yanzhen and You, Jiaxuan},
  journal={arXiv preprint arXiv:2410.03834},
  year={2024}
}

@article{chen2024routerdc,
  title={Routerdc: Query-based router by dual contrastive learning for assembling large language models},
  author={Chen, Shuhao and Jiang, Weisen and Lin, Baijiong and Kwok, James and Zhang, Yu},
  journal={Advances in Neural Information Processing Systems},
  volume={37},
  pages={66305--66328},
  year={2024}
}

@article{mei2024aios,
  title={AIOS: LLM agent operating system},
  author={Mei, Kai and Li, Zelong and Xu, Shuyuan and Ye, Ruosong and Ge, Yingqiang and Zhang, Yongfeng},
  journal={arXiv e-prints, pp. arXiv--2403},
  year={2024}
}

@article{lewis2020retrieval,
  title={Retrieval-augmented generation for knowledge-intensive nlp tasks},
  author={Lewis, Patrick and Perez, Ethan and Piktus, Aleksandra and Petroni, Fabio and Karpukhin, Vladimir and Goyal, Naman and K{\"u}ttler, Heinrich and Lewis, Mike and Yih, Wen-tau and Rockt{\"a}schel, Tim and others},
  journal={Advances in Neural Information Processing Systems},
  volume={33},
  pages={9459--9474},
  year={2020}
}

@article{packer2023memgpt,
  title={Memgpt: Towards llms as operating systems},
  author={Packer, Charles and Fang, Vivian and Patil, Shishir G and Lin, Kevin and Wooders, Sarah and Gonzalez, Joseph E},
  journal={arXiv preprint arXiv:2310.08560},
  year={2023}
}

@article{jin2023s3,
  title={S3: Increasing GPU Utilization during Generative Inference for Higher Throughput},
  author={Jin, Yunho and Wu, Chun-Feng and Brooks, David and Wei, Gu-Yeon},
  journal={Advances in Neural Information Processing Systems},
  volume={36},
  pages={18015--18027},
  year={2023}
}

@inproceedings{zheng2024response,
  title={Response length perception and sequence scheduling: An llm-empowered llm inference pipeline},
  author={Zheng, Zangwei and Ren, Xiaozhe and Xue, Fuzhao and Luo, Yang and Jiang, Xin and You, Yang},
  journal={Advances in Neural Information Processing Systems},
  volume={36},
  year={2024}
}

@inproceedings{aiops2024qiu,
  author  = {Qiu, Haoran and Mao, Weichao and Patke, Archit and Cui, Shengkun and Jha, Saurabh and Wang, Chen and Franke, Hubertus and Kalbarczyk, Zbigniew T. and Ba\c{s}ar, Tamer and Iyer, Ravishankar K.},
  title   = {Efficient Interactive LLM Serving with Proxy Model-based Sequence Length Prediction},
  year    = {2024},
  pages = {1--7},
  publisher = {Association for Computing Machinery},
  volume = {5},
  address = {San Diego, CA, USA},
  booktitle = {The 5th International Workshop on Cloud Intelligence / AIOps at ASPLOS 2024},
}

@article{cheng2024enabling,
  title={Enabling Efficient Batch Serving for LMaaS via Generation Length Prediction},
  author={Cheng, Ke and Hu, Wen and Wang, Zhi and Du, Peng and Li, Jianguo and Zhang, Sheng},
  journal={arXiv preprint arXiv:2406.04785},
  year={2024}
}

@inproceedings{qiu2024power,
  title={Power-aware Deep Learning Model Serving with $\{$$\mu$-Serve$\}$},
  author={Qiu, Haoran and Mao, Weichao and Patke, Archit and Cui, Shengkun and Jha, Saurabh and Wang, Chen and Franke, Hubertus and Kalbarczyk, Zbigniew and Ba{\c{s}}ar, Tamer and Iyer, Ravishankar K},
  booktitle={2024 USENIX Annual Technical Conference (USENIX ATC 24)},
  pages={75--93},
  year={2024}
}

@article{cobbe2021training,
  title={Training verifiers to solve math word problems},
  author={Cobbe, Karl and Kosaraju, Vineet and Bavarian, Mohammad and Chen, Mark and Jun, Heewoo and Kaiser, Lukasz and Plappert, Matthias and Tworek, Jerry and Hilton, Jacob and Nakano, Reiichiro and others},
  journal={arXiv preprint arXiv:2110.14168},
  year={2021}
}

@article{hendrycks2021measuring,
  title={Measuring mathematical problem solving with the math dataset},
  author={Hendrycks, Dan and Burns, Collin and Kadavath, Saurav and Arora, Akul and Basart, Steven and Tang, Eric and Song, Dawn and Steinhardt, Jacob},
  journal={arXiv preprint arXiv:2103.03874},
  year={2021}
}

@article{fu2024efficient,
  title={Efficient LLM Scheduling by Learning to Rank},
  author={Fu, Yichao and Zhu, Siqi and Su, Runlong and Qiao, Aurick and Stoica, Ion and Zhang, Hao},
  journal={arXiv preprint arXiv:2408.15792},
  year={2024}
}

@article{hendrycks2020measuring,
  title={Measuring massive multitask language understanding},
  author={Hendrycks, Dan and Burns, Collin and Basart, Steven and Zou, Andy and Mazeika, Mantas and Song, Dawn and Steinhardt, Jacob},
  journal={arXiv preprint arXiv:2009.03300},
  year={2020}
}

@inproceedings{zhang2020solving,
  title={Solving billion-scale knapsack problems},
  author={Zhang, Xingwen and Qi, Feng and Hua, Zhigang and Yang, Shuang},
  booktitle={Proceedings of The Web Conference 2020},
  pages={3105--3111},
  year={2020}
}

@book{bertsekas2014constrained,
  title={Constrained optimization and Lagrange multiplier methods},
  author={Bertsekas, Dimitri P},
  year={2014},
  publisher={Academic press}
}

@article{hui2024qwen2,
  title={Qwen2. 5-coder technical report},
  author={Hui, Binyuan and Yang, Jian and Cui, Zeyu and Yang, Jiaxi and Liu, Dayiheng and Zhang, Lei and Liu, Tianyu and Zhang, Jiajun and Yu, Bowen and Lu, Keming and others},
  journal={arXiv preprint arXiv:2409.12186},
  year={2024}
}

@article{homaifar1994constrained,
  title={Constrained optimization via genetic algorithms},
  author={Homaifar, Abdollah and Qi, Charlene X and Lai, Steven H},
  journal={Simulation},
  volume={62},
  number={4},
  pages={242--253},
  year={1994},
  publisher={Sage Publications Sage CA: Thousand Oaks, CA}
}

@inproceedings{shi2025from,
  title={From Commands to Prompts: {LLM}-based Semantic File System for AIOS},
  author={Zeru Shi and Kai Mei and Mingyu Jin and Yongye Su and Chaoji Zuo and Wenyue Hua and Wujiang Xu and Yujie Ren and Zirui Liu and Mengnan Du and Dong Deng and Yongfeng Zhang},
  booktitle={The Thirteenth International Conference on Learning Representations},
  year={2025},
  url={https://openreview.net/forum?id=2G021ZqUEZ}
}

@article{zheng2023efficiently,
  title={Efficiently Programming Large Language Models using SGLang.},
  author={Zheng, Lianmin and Yin, Liangsheng and Xie, Zhiqiang and Huang, Jeff and Sun, Chuyue and Yu, Cody\_Hao and Cao, Shiyi and Kozyrakis, Christos and Stoica, Ion and Gonzalez, Joseph E and others},
  year={2023},
  publisher={arXiv}
}

@inproceedings{kwon2023efficient,
  title={Efficient memory management for large language model serving with pagedattention},
  author={Kwon, Woosuk and Li, Zhuohan and Zhuang, Siyuan and Sheng, Ying and Zheng, Lianmin and Yu, Cody Hao and Gonzalez, Joseph and Zhang, Hao and Stoica, Ion},
  booktitle={Proceedings of the 29th Symposium on Operating Systems Principles},
  pages={611--626},
  year={2023}
}

@article{zhang2024ai,
  title={When ai meets finance (stockagent): Large language model-based stock trading in simulated real-world environments},
  author={Zhang, Chong and Liu, Xinyi and Zhang, Zhongmou and Jin, Mingyu and Li, Lingyao and Wang, Zhenting and Hua, Wenyue and Shu, Dong and Zhu, Suiyuan and Jin, Xiaobo and others},
  journal={arXiv preprint arXiv:2407.18957},
  year={2024}
}

@article{hua2024disentangling,
  title={Disentangling logic: The role of context in large language model reasoning capabilities},
  author={Hua, Wenyue and Zhu, Kaijie and Li, Lingyao and Fan, Lizhou and Lin, Shuhang and Jin, Mingyu and Xue, Haochen and Li, Zelong and Wang, JinDong and Zhang, Yongfeng},
  journal={arXiv preprint arXiv:2406.02787},
  year={2024}
}

@inproceedings{hua2024trustagent,
  title={TrustAgent: Towards Safe and Trustworthy LLM-based Agents},
  author={Hua, Wenyue and Yang, Xianjun and Jin, Mingyu and Li, Zelong and Cheng, Wei and Tang, Ruixiang and Zhang, Yongfeng},
  booktitle={2024 Conference on Empirical Methods in Natural Language Processing, EMNLP 2024},
  pages={10000--10016},
  year={2024},
  organization={Association for Computational Linguistics (ACL)}
}

@inproceedings{wang2023multi,
  title={A Multi-stage Framework for Online Bonus Allocation Based on Constrained User Intent Detection},
  author={Wang, Chao and Shi, Xiaowei and Xu, Shuai and Wang, Zhe and Fan, Zhiqiang and Feng, Yan and You, An and Chen, Yu},
  booktitle={Proceedings of the 29th ACM SIGKDD Conference on Knowledge Discovery and Data Mining},
  pages={5028--5038},
  year={2023}
}

@article{zhu2024deepseek,
  title={DeepSeek-Coder-V2: Breaking the Barrier of Closed-Source Models in Code Intelligence},
  author={Zhu, Qihao and Guo, Daya and Shao, Zhihong and Yang, Dejian and Wang, Peiyi and Xu, Runxin and Wu, Y and Li, Yukun and Gao, Huazuo and Ma, Shirong and others},
  journal={arXiv preprint arXiv:2406.11931},
  year={2024}
}

@article{devlin2018bert,
  title={Bert: Pre-training of deep bidirectional transformers for language understanding},
  author={Devlin, Jacob},
  journal={arXiv preprint arXiv:1810.04805},
  year={2018}
}

@article{wei2023magicoder,
  title={Magicoder: Source code is all you need},
  author={Wei, Yuxiang and Wang, Zhe and Liu, Jiawei and Ding, Yifeng and Zhang, Lingming},
  journal={arXiv preprint arXiv:2312.02120},
  year={2023}
}

@article{nijkamp2023codegen2,
  title={Codegen2: Lessons for training llms on programming and natural languages},
  author={Nijkamp, Erik and Hayashi, Hiroaki and Xiong, Caiming and Savarese, Silvio and Zhou, Yingbo},
  journal={arXiv preprint arXiv:2305.02309},
  year={2023}
}

@article{sun2024llumnix,
  title={Llumnix: Dynamic Scheduling for Large Language Model Serving},
  author={Sun, Biao and Huang, Ziming and Zhao, Hanyu and Xiao, Wencong and Zhang, Xinyi and Li, Yong and Lin, Wei},
  journal={arXiv preprint arXiv:2406.03243},
  year={2024}
}

@article{rein2023gpqa,
  title={Gpqa: A graduate-level google-proof q\&a benchmark},
  author={Rein, David and Hou, Betty Li and Stickland, Asa Cooper and Petty, Jackson and Pang, Richard Yuanzhe and Dirani, Julien and Michael, Julian and Bowman, Samuel R},
  journal={arXiv preprint arXiv:2311.12022},
  year={2023}
}

@article{hu2024inference,
  title={Inference without interference: Disaggregate llm inference for mixed downstream workloads},
  author={Hu, Cunchen and Huang, Heyang and Xu, Liangliang and Chen, Xusheng and Xu, Jiang and Chen, Shuang and Feng, Hao and Wang, Chenxi and Wang, Sa and Bao, Yungang and others},
  journal={arXiv preprint arXiv:2401.11181},
  year={2024}
}

@article{stojkovic2024dynamollm,
  title={Dynamollm: Designing llm inference clusters for performance and energy efficiency},
  author={Stojkovic, Jovan and Zhang, Chaojie and Goiri, {\'I}{\~n}igo and Torrellas, Josep and Choukse, Esha},
  journal={arXiv preprint arXiv:2408.00741},
  year={2024}
}

@article{achiam2023gpt,
  title={Gpt-4 Technical Report},
  author={Achiam, Josh and Adler, Steven and Agarwal, Sandhini and Ahmad, Lama and Akkaya, Ilge and Aleman, Florencia Leoni and Almeida, Diogo and Altenschmidt, Janko and Altman, Sam and Anadkat, Shyamal and others},
  journal={arXiv preprint:2303.08774 (OpenAI Technical Report)},
  year={2023}
}

@article{team2023gemini,
  title={Gemini: a family of highly capable multimodal models},
  author={Team, Gemini and Anil, Rohan and Borgeaud, Sebastian and Alayrac, Jean-Baptiste and Yu, Jiahui and Soricut, Radu and Schalkwyk, Johan and Dai, Andrew M and Hauth, Anja and Millican, Katie and others},
  journal={arXiv preprint arXiv:2312.11805},
  year={2023}
}

@article{guo2025deepseek,
  title={Deepseek-r1: Incentivizing reasoning capability in llms via reinforcement learning},
  author={Guo, Daya and Yang, Dejian and Zhang, Haowei and Song, Junxiao and Zhang, Ruoyu and Xu, Runxin and Zhu, Qihao and Ma, Shirong and Wang, Peiyi and Bi, Xiao and others},
  journal={arXiv preprint arXiv:2501.12948},
  year={2025}
}

@article{dubey2024llama,
  title={The llama 3 herd of models},
  author={Dubey, Abhimanyu and Jauhri, Abhinav and Pandey, Abhinav and Kadian, Abhishek and Al-Dahle, Ahmad and Letman, Aiesha and Mathur, Akhil and Schelten, Alan and Yang, Amy and Fan, Angela and others},
  journal={arXiv preprint arXiv:2407.21783 (Meta AI Technical Report)},
  year={2024}
}

@article{li2024llms,
  title={Llms-as-judges: a comprehensive survey on llm-based evaluation methods},
  author={Li, Haitao and Dong, Qian and Chen, Junjie and Su, Huixue and Zhou, Yujia and Ai, Qingyao and Ye, Ziyi and Liu, Yiqun},
  journal={arXiv preprint arXiv:2412.05579},
  year={2024}
}

@article{guo2025deepsieve,
  title={DeepSieve: Information Sieving via LLM-as-a-Knowledge-Router},
  author={Guo, Minghao and Zeng, Qingcheng and Zhao, Xujiang and Liu, Yanchi and Yu, Wenchao and Du, Mengnan and Chen, Haifeng and Cheng, Wei},
  journal={arXiv preprint arXiv:2507.22050},
  year={2025}
}

@article{amem,
  title={A-mem: Agentic memory for llm agents},
  author={Xu, Wujiang and Liang, Zujie and Mei, Kai and Gao, Hang and Tan, Juntao and Zhang, Yongfeng},
  journal={arXiv preprint arXiv:2502.12110},
  year={2025}
}

@article{iagent,
  title={iAgent: LLM Agent as a Shield between User and Recommender Systems},
  author={Xu, Wujiang and Shi, Yunxiao and Liang, Zujie and Ning, Xuying and Mei, Kai and Wang, Kun and Zhu, Xi and Xu, Min and Zhang, Yongfeng},
  journal={arXiv preprint arXiv:2502.14662},
  year={2025}
}

\end{document}